\PassOptionsToPackage{numbers,square,comma}{ai2style/natbib}

\documentclass[table]{ai2style/ai2}

\usepackage[utf8]{inputenc}
\usepackage[T1]{fontenc}
\usepackage{hyperref}
\usepackage{url}
\usepackage{graphicx}
\usepackage{booktabs}
\usepackage{amsmath,amssymb}
\usepackage{enumitem}
\usepackage{array}
\usepackage{tabularx}
\usepackage{longtable}
\usepackage{makecell}
\usepackage{multirow}
\usepackage{caption}
\usepackage{subcaption}
\usepackage{xspace}
\usepackage{microtype}
\usepackage{threeparttable}
\usepackage{float}
\usepackage{placeins}
\usepackage{pifont}

\definecolor{darkblue}{rgb}{0,0,0.5}
\hypersetup{colorlinks=true, citecolor=darkblue, linkcolor=darkblue, urlcolor=darkblue}

\setlist[itemize]{leftmargin=*,itemsep=0em,parsep=0.3em,topsep=0.3em}
\setlist[enumerate]{leftmargin=*,itemsep=0em,parsep=0.3em,topsep=0.3em}
\DeclareUnicodeCharacter{2212}{\ensuremath{-}}

\newcommand{\jdhAlgoAff}{\raisebox{.28em}{\hspace{.02em}\scalebox{0.7}{\textbf{1}}}}
\newcommand{\commaAff}{\raisebox{.28em}{\hspace{.02em}\scalebox{0.7}{\textbf{,}\hspace{0.1em}}}}

\newcommand{\hospitalAff}{\raisebox{.28em}{\hspace{.02em}\scalebox{0.7}{\textbf{2}}}}
\newcommand{\centerAff}{\raisebox{.28em}{\hspace{.02em}\scalebox{0.7}{\textbf{3}}}}
\newcommand{\icuAff}{\raisebox{.28em}{\hspace{.02em}\scalebox{0.7}{\textbf{4}}}}

\renewcommand\authorOne[2][]{\addtolist[#1]{#2}{\authorlistOne}{\authorformat}{\hspace{0.72em}}}
\renewcommand\authorformat[2][]{\mbox{\small\sffamily \bfseries #2#1}}
\renewcommand\affiliation[2][]{\addtolist[#1]{#2}{\affiliationlist}{\affiliationformat}{\par}}
\title{RadSEM: A Finding-by-Finding Metric for Clinical Consistency in Radiology Reports}

\authorOne[\jdhAlgoAff]{Zhenhong Yang\textsuperscript{*}}
\authorOne[\jdhAlgoAff]{Zhuoyun Liu\textsuperscript{*}}
\authorOne[\jdhAlgoAff]{Jintao Fei}
\authorOne[\jdhAlgoAff]{Wen Tang}
\authorOne[\hospitalAff\commaAff\centerAff\commaAff\icuAff]{Shichao Quan\textsuperscript{\textdagger}}
\authorOne[\jdhAlgoAff]{Jun Zhao\textsuperscript{\textdagger}}
\authorOne[\jdhAlgoAff]{Jun Xu}

\affiliation[\jdhAlgoAff\,]{JDH Algo, JD Health International Inc., China} 
\affiliation[\hospitalAff\,]{Department of Big Data in Health Science, The First Affiliated Hospital of Wenzhou Medical University, China}
\affiliation[\centerAff\,]{Zhejiang Engineering Research Center for Hospital Emergency and Process Digitization, Wenzhou, China}
\affiliation[\icuAff\,]{Department of Intensive Care Unit, The First Affiliated Hospital of Wenzhou Medical University, Wenzhou, Zhejiang, China}

\contribution[*]{Equal contribution}
\contribution[\dagger]{Corresponding authors.}

\metadata[Correspondence:]{\href{mailto:zhaojun10@jd.com}{\color{ai2accent}{\texttt{zhaojun10@jd.com}}}}

\abstract{
Radiology reports are built from compact clinical statements, where a single negation, laterality marker, or normal--abnormal polarity can change the clinical meaning. Existing metrics usually approach report evaluation from one of three directions: they compare textual resemblance, make holistic error judgments, or align extracted fine-grained findings. These views are useful, but they do not always systematically answer a local clinical question: does each clinical fact in a generated report agree with the reference facts, especially when abnormalities are missed or altered, or when statements that cannot be true together are nevertheless treated as similar? We propose RadSEM (Radiology Sentence-Level Evaluation Metric), a constrained LLM-assisted metric for reference-based scoring of radiology Findings. RadSEM first rewrites reference and generated reports into ordered \emph{atomic finding sentences}, each expressing one site--finding proposition after non-finding content has been removed. It then performs contradiction-constrained many-to-many semantic matching between these units, so that lexically similar but clinically incompatible statements, such as ``no effusion'' and ``effusion,'' do not receive credit, while clinically compatible granularity differences can still be represented as pair-level partial matches. Each accepted pair is weighted according to part--whole relationships and abnormal-detail coverage, and a deterministic scoring stage aggregates the pair weights under one unit of matched-credit capacity per atomic sentence, together with unmatched sentences, into an abnormal-focused weighted F1 score. In this design, the LLM supports structured rewriting and local semantic alignment rather than acting as an opaque global judge. RadSEM focuses on Findings rather than Impressions to reduce confounding between image-reading agreement and context-dependent diagnostic synthesis. We assess RadSEM with SSREE, a controlled monotonicity stress test built from 2,448 de-identified, English-translated real reports and five corruption levels (12,240 scoring instances), and with a 599-triplet synonym/antonym rewrite subset. In these controlled experiments, RadSEM obtains Kendall $\tau_b$ of 0.957, all-pairs concordance of 97.8\%, adjacent-level concordance of 95.0\%, and strict five-level ordering for 81.9\% of reports on SSREE; on the rewrite subset, it prefers synonymous over antonymous rewrites in 597 of 599 cases (99.67\%). These results suggest that explicit statement units and contradiction-constrained matching can make automatic report scoring more interpretable and less vulnerable to clinically misleading surface similarity. The code is available at \href{https://github.com/jdh-algo/RadSEM}{https://github.com/jdh-algo/RadSEM}.
}

\usepackage{xspace}

\newif\ifsinglecolumn

\definecolor{darkgreen}{RGB}{0, 100, 0}
\definecolor{darkred}{RGB}{139, 0, 0}

\begin{document}
\maketitle

\section{Introduction}

Radiology reports are compact collections of clinical assertions, where a small semantic operator can carry most of the clinical meaning. A single change may invert a finding: ``no effusion'' versus ``effusion,'' left versus right, normal versus abnormal, or decreased versus increased. As radiology-report generation models are increasingly developed and evaluated for summarizing imaging findings, automatic scoring should therefore ask more than whether a generated report sounds fluent or resembles a reference. It should ask whether the individual observations it states are clinically compatible with the observations in the reference.

This distinction is often blurred by existing automatic metrics. Generic overlap- and embedding-based scores may reward shared anatomy terms, disease words, or surrounding context even when the candidate contradicts the reference. Radiology-specific metrics improve on generic text scores by modeling clinical entities, relations, error types, or fine-grained finding labels, and they provide important precedents for structured report evaluation. Yet practical scoring can still be affected by extractor coverage, modality-specific schemas, granularity mismatch, and overly permissive alignment. The underlying problem is that semantic relatedness is not the same as clinical compatibility: two sentences can be close in wording and topic while describing states that cannot both be true for the same patient, site, and time.

Two issues recur in many commonly used scoring settings. First, \textbf{evaluation units can be unstable}. Whole-report comparison mixes multiple observations, raw sentence comparison depends on writing style, and coarse entity sets may separate modifiers from the finding they qualify. One radiologist may express several observations in a single sentence, while another may describe the same content as several shorter statements. Prior fine-grained finding-label representations address part of this problem by normalizing findings and modifiers; RadSEM follows the same broad principle but targets reference-based free-text scoring with inspectable natural-language units. Second, \textbf{matching is often unconstrained}. A soft alignment procedure may pair contradictory statements and still assign partial credit because they share anatomy, disease terms, or lexical context. Together, unstable units and unconstrained matching make report-level or raw sentence-level similarity an unreliable proxy for observation-level correctness.

RadSEM evaluates the Findings section because its intended target is image-reading fidelity. In routine clinical practice, an Impression is a diagnosis-oriented synthesis that may combine imaging observations with patient history, current clinical status, prior information, and the question posed by the referring clinician. When an in-hospital report is used as a reference for scoring an image-reading model, the Impression can therefore become a confounding target: it may contain context-dependent conclusions that are not recoverable from the image alone. RadSEM does not score the Impression section and instead focuses on observation-level agreement in Findings, rather than claiming to assess the full quality of diagnostic synthesis.

RadSEM, short for Radiology Sentence-Level Evaluation Metric, addresses the above issues by aligning the comparison unit with how many report discrepancies are clinically reviewed: as individual statements about a site and a finding. In this name, sentence-level refers to the rewritten atomic finding sentence used for scoring, not to raw sentence boundaries in the original report. It first rewrites both the reference report (\textbf{Ref}) and generated report (\textbf{Gen}) into ordered lists of \emph{atomic finding sentences}, each expressing one clinical proposition in a structured natural-language form. It then matches these statements under an explicit anti-contradiction constraint: if two statements cannot be simultaneously true, the pair is forbidden and receives no credit. Because independently rewritten reports may express the same clinical content at different granularities, matching is many-to-many at the pair level: broader and more specific atomic statements can be linked in either direction when the clinical meaning is compatible. Accepted pairs receive partial credit for clinically defensible differences, such as part--whole granularity or incomplete abnormal-detail coverage. Finally, deterministic code aggregates the emitted pair weights under sentence-level matched-credit capacities and combines them with unmatched statements into an abnormal-focused weighted F1 score that emphasizes abnormal findings while still representing normal-background statements.

Our contributions are threefold:
\begin{itemize}
    \item We propose RadSEM, a constrained LLM-assisted scoring framework that treats radiology report scoring as clinically inspectable statement comparison rather than global text similarity. RadSEM rewrites free-text Findings into ordered finding units for reference-based scoring while complementing prior structured representations.
    \item We introduce contradiction-constrained many-to-many semantic matching, which separates semantic relatedness from clinical compatibility. By exposing clinically compatible many-to-many pair links, forbidding alignments between statements that cannot be simultaneously true, assigning partial credit for defensible granularity or detail-coverage gaps, aggregating matched credit under sentence-level capacities, and explicitly accounting for unmatched Ref/Gen sentences, RadSEM makes the local logic of credit assignment inspectable.
    \item We design and apply SSREE, a controlled monotonicity stress test that evaluates whether metrics respond consistently to graded semantic corruption. On 2,448 de-identified, English-translated real reports expanded into five corruption levels, RadSEM shows higher ranking consistency than the evaluated radiology-specific and general-purpose baselines.
\end{itemize}

\section{Related Work}

Automatic evaluation for radiology report generation can be organized by the structure that a metric exposes before assigning a score. Some methods compare extracted clinical entities or relations; others map reports into fine-grained finding labels, ask an LLM to act as a clinical reviewer, or use generic text similarity as a lightweight proxy. We review these directions with particular attention to the unit being scored and the degree to which the scoring policy is explicit.

\subsection{Structured radiology-report metrics}

\paragraph{Entity-/relation-based overlap: RadGraph F1.}
A common strategy is to transform free text into a structured representation and then compute overlap in the extracted clinical content. RadGraph provides an entity--relation information extraction schema and benchmark for radiology reports, enabling conversion from sentences to a clinical graph representation \citep{jain2021radgraph}. Building on this structure, RadGraph F1 evaluates a candidate report by comparing extracted RadGraph entities and relations against those from the reference. This structure is intended to make the score more sensitive to clinically meaningful content differences than surface n-gram metrics. This direction is empirically motivated by studies that analyze correlations with radiologist judgments and propose RadGraph F1 and composites such as RadCliQ as better-aligned automatic metrics for chest X-ray report generation \citep{yu2023progress}. A known limitation of this family is that the score is bounded by the coverage and accuracy of the underlying extractor, including schema granularity, parser errors, and domain shifts.

\paragraph{Entity-aware semantic scoring: RaTEScore.}
Another line of work combines medical named-entity recognition (NER) with semantic similarity to retain domain awareness while allowing more linguistic variation. RaTEScore decomposes radiology reports into medically salient entities and then compares entity representations with type- and importance-aware weighting, emphasizing diagnostic outcomes and anatomical details while being sensitive to negations and robust to synonyms \citep{zhao2024ratescore}. Compared with pure concept-label overlap, entity-aware semantic scoring can reduce brittleness to paraphrasing, but it still depends on the quality of the NER model and the entity schema. Its score is also less directly auditable than an atomic finding sentence alignment, because entity similarity and weighting are aggregated without always exposing which clinical statement was accepted, partially accepted, or rejected. This also illustrates a broader difficulty in radiology metric validation: human error counts and preference judgments can depend on segmentation, scoring granularity, and aggregation conventions. This motivates complementary stress tests that use controlled semantic perturbations and predefined quality orderings rather than relying only on subjective error counts.

\paragraph{Fine-grained finding labels and phrasal grounding.}
A closely related line of work represents chest radiology reports through fine-grained finding labels (FFL), which encode a core finding together with presence or absence and modifiers such as anatomical location, laterality, severity, size, and appearance. Early FFL work extracted such labels from reports by combining vocabulary-driven concept extraction, dependency-based phrasal grouping, negation detection, and pattern completion, and then used the labels for fine-grained chest X-ray label learning and report generation \citep{syedamahmood2020extracting,syedamahmood2020chest}. More recent work extends this representation to report evaluation by extracting fine-grained finding patterns from both ground-truth and generated reports, computing textual FFL precision/recall/F1, and combining text-level matching with phrasal grounding over anatomical regions, including Chest ImaGenome-style anatomical boxes \citep{mahmood2024evaluating,wu2021chestimagenome}. This line of work is especially important for RadSEM because it already addresses several problems that motivate our metric, including finding-level decomposition, modifier-aware comparison, and anatomy-aware grounding. RadSEM should therefore not be interpreted as the first attempt to structure radiology reports into fine-grained clinical units. RadSEM differs by targeting reference-based free-text evaluation without requiring a closed chest-X-ray label vocabulary or image-region annotations, while making contradiction handling, partial credit, and abnormal-focused aggregation explicit.

\paragraph{LLM-based error auditing: GREEN.}
Large-language-model-based evaluators have also been explored to judge clinically significant errors and provide explanations. GREEN (Generative Radiology Report Evaluation and Error Notation) uses language-model understanding to identify and explain clinically important mistakes, producing both a scalar score and interpretable error rationales \citep{ostmeier2024green}. This makes GREEN valuable as an expert-style reviewer: it can point to false findings, missed findings, wrong locations, incorrect severity, or other clinically significant discrepancies in natural language. At the same time, this flexibility means that the metric-level policy for counting findings, deciding partial matches, and separating harmless refinements from unsupported details depends on the judge model, prompt, and implicit reviewer convention. GREEN is therefore highly useful for error auditing and human-readable analysis, whereas a reproducible benchmark metric may benefit from stricter control over the units, labels, and aggregation rules that produce the final score.

\subsection{Platform-level and general-purpose evaluation}

\paragraph{Platform-level key-point scoring and LLM judging.}
Beyond radiology-specific metrics, medical benchmark platforms increasingly evaluate heterogeneous tasks through structured scoring interfaces and LLM-as-a-judge protocols. MedBench v4, for example, is designed for large-scale evaluation of medical language models, multimodal models, and agents, with open-ended responses scored by an LLM judge calibrated to human ratings \citep{ding2025medbenchv4}. In report-generation settings, macro-recall over predefined key information points, optionally complemented by judge-based assessment, offers a practical platform-level interface: key points make coverage measurable, and the judge can account for broader aspects such as correctness, professional language, safety, and usability. This design is useful for scalable leaderboard-style evaluation, but it primarily standardizes how responses are scored rather than defining radiology-specific clinical content units. The score can depend on how reference information is segmented into key points, and recall-centered aggregation alone does not penalize unsupported extra findings unless the judge component captures them. MedBench-based analyses likewise treat omissions and hallucinations as consequential failure modes for medical LLM evaluation \citep{jiang2025benchmarking}.

\paragraph{N-gram overlap metrics: BLEU, ROUGE, and CIDEr.}
General-purpose metrics remain widely used due to simplicity, speed, and historical comparability across generation tasks. BLEU computes modified n-gram precision with a brevity penalty \citep{papineni2002bleu}. ROUGE, including ROUGE-N and ROUGE-L, emphasizes recall-oriented overlap and was originally developed for summarization evaluation \citep{lin2004rouge}. CIDEr weights n-gram overlap by TF--IDF to emphasize informative phrases and is common in captioning-style generation \citep{vedantam2015cider}. In radiology, these metrics can be misleading because high token overlap may persist despite clinically decisive flips such as negation, laterality, or normal/abnormal polarity, and because they do not enforce contradiction awareness.

\paragraph{Soft alignment metrics: METEOR.}
METEOR computes a unigram matching score with explicit alignment and heuristic penalties, aiming for stronger correlation with human judgments in machine translation than raw n-gram precision alone \citep{banerjee2005meteor}. While this alignment is less brittle to minor lexical variation, it is not designed to represent clinical truth conditions. In particular, it does not explicitly model medical entities, relations, or contradiction constraints.

\subsection{Design space and positioning of RadSEM}

The above methods occupy different points in the design space of report evaluation. RadGraph and RaTEScore introduce clinical structure and semantic similarity, but their scoring units are determined by the extractor or entity schema. FFL and phrasal-grounding methods make the clinical units more explicit through normalized finding patterns and, when image-region annotations are available, anatomical grounding. GREEN emphasizes expert-style LLM review and human-interpretable error explanations. Platform-level key-point and LLM-judge protocols emphasize scalable comparison across heterogeneous medical tasks. Generic text metrics provide inexpensive baselines but do not represent clinical truth conditions.

RadSEM is positioned as a constrained LLM-assisted atomic finding sentence metric within this landscape. It does not claim to be the first method to decompose radiology reports into fine-grained finding units, and it does not perform image-grounded fact checking. Its distinction is narrower and more operational: compared with RadGraph-style metrics, it does not depend on a fixed entity--relation extractor as the scored object; compared with RaTEScore, it exposes statement-level matches and rejections rather than only aggregated entity similarity; compared with FFL and phrasal-grounding methods, it targets open free-text Findings without requiring a closed chest-X-ray label vocabulary or image-region annotations; and compared with GREEN, it avoids asking the LLM for a free-form overall judgment. Instead, RadSEM converts free-text reports into inspectable finding units, constrains semantic matching with anti-contradiction rules, assigns partial-credit labels, and computes an abnormal-focused weighted F1 score through explicit aggregation rules. Thus, the LLM helps expose candidate clinical statements and semantic alignments, while the scoring policy remains reproducible once those structured tags are fixed.

\section{RadSEM}
\label{sec:method}

\begin{figure}[!t]
\centering
\includegraphics[width=0.95\textwidth]{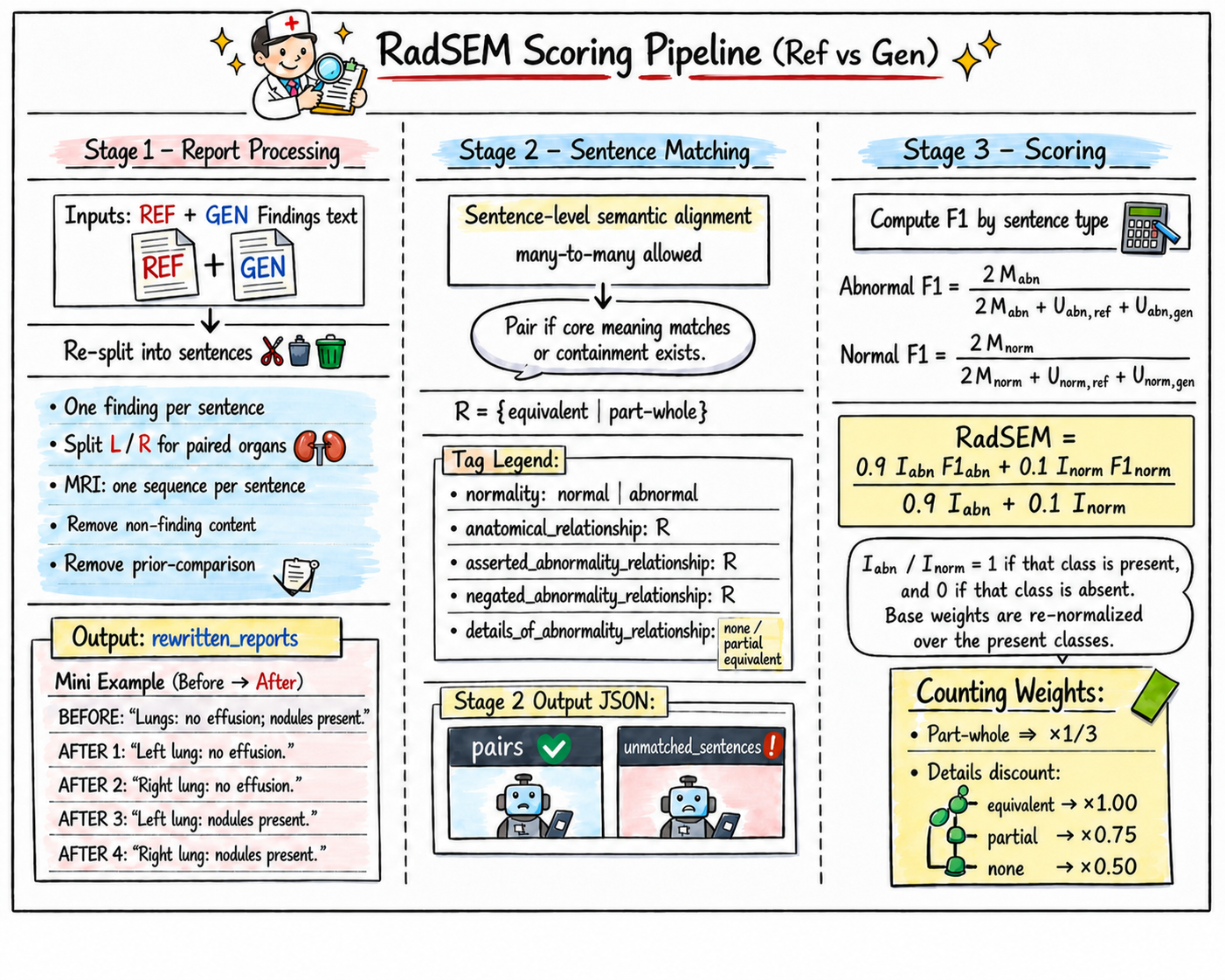}
\caption{RadSEM pipeline. Stage 1 (report processing) rewrites Ref/Gen into ordered atomic finding sentences; Stage 2 (sentence matching) identifies clinically compatible sentence pairs with an explicit anti-contradiction constraint and allows many-to-many pair alignment when clinical meaning is compatible across different granularities; Stage 3 (scoring) computes an abnormal-focused weighted F1 score using pair-level relationship labels, sentence-level matched-credit capacities, and unmatched Ref/Gen sentences.}
\label{fig:radsem-pipeline}
\end{figure}

RadSEM evaluates a generated radiology report (\textbf{Gen}) against a reference report (\textbf{Ref}) through a constrained LLM-assisted pipeline followed by deterministic scoring. Its atomic finding sentence representation is conceptually related to prior structured finding-level representations, including entity--relation graphs and fine-grained finding labels, which also decompose reports into clinically meaningful units with modifiers. RadSEM differs in implementation and scoring: it retains natural-language atomic finding sentences rather than forcing every statement into a closed chest-X-ray label vocabulary, and it separates LLM-assisted semantic parsing from deterministic aggregation. The released implementation follows the same three stages shown in Figure~\ref{fig:radsem-pipeline}: Stage 1 (report processing) rewrites both Gen and Ref into atomic finding sentences, Stage 2 (sentence matching) identifies clinically compatible Ref--Gen sentence pairs under a many-to-many alignment policy, and Stage 3 (scoring) aggregates the structured pair and unmatched-sentence tags into a scalar abnormal-focused score using capacity-constrained matched credit. Let $A^{\mathrm{Ref}}=(a_1,\dots,a_m)$ and $A^{\mathrm{Gen}}=(b_1,\dots,b_n)$ denote the ordered atomic finding sentence lists produced from Ref and Gen.

\subsection{Pipeline responsibilities}

RadSEM is designed so that the language model does not directly assign a free-form overall quality score. Instead, the LLM is restricted to two local tasks: rewriting reports into comparable clinical units and labeling semantic alignments between those units. The implementation uses fixed task prompts for these two constrained LLM stages; intermediate structured outputs can be inspected so that disagreements caused by rewriting, matching, or scoring are separable. The final score is produced by fixed aggregation code rather than by the LLM. Table~\ref{tab:radsem-stage-responsibilities} summarizes this division of labor. The pipeline rewrites generated reports and reference reports independently with the same rules, aligns samples by case name, produces a structured tag file containing accepted sentence pairs, pair labels, and unmatched atomic finding sentences, and then writes per-case scores. This symmetric preprocessing is important: both sides are converted to the same type of clinical unit before any comparison is made.

\begin{table}[!t]
\centering
\caption{Responsibilities of the three RadSEM stages. The first two stages use constrained LLM-assisted semantic processing; the final stage is deterministic given the structured tags.}
\label{tab:radsem-stage-responsibilities}
\small
\begin{tabularx}{\textwidth}{p{0.17\textwidth}p{0.37\textwidth}p{0.38\textwidth}}
\toprule
\textbf{Stage} & \textbf{Responsibility} & \textbf{Scoring role} \\
\midrule
\makecell[l]{Stage 1:\\Report processing} & Rewrite only the supplied Findings text, split compound descriptions, isolate laterality and sequence-specific observations when needed, remove non-finding content and comparison-to-prior wording, remove identical duplicate sentences, and keep the original finding order. & Standardizes report prose into ordered atomic finding sentences without assigning credit or comparing reports. \\
\addlinespace
\makecell[l]{Stage 2:\\Sentence matching} & Compare rewritten Ref and Gen reports aligned by case name; identify clinically compatible Ref--Gen sentence pairs under a many-to-many alignment policy; apply anti-contradiction rules; label normality, anatomy, asserted/negated abnormality concepts, abnormal-detail coverage, and unmatched sentences. & Exposes the local semantic evidence used for omissions, unsupported additions, partial matches, and contradictions. \\
\addlinespace
\makecell[l]{Stage 3:\\Scoring} & Allocate weighted matched credit separately for abnormal and normal classes subject to one unit of capacity for each Ref/Gen atomic finding sentence, count unmatched Ref/Gen atomic finding sentences, apply the deterministic no-unmatched class rule and no-unmatched partial adjustment, and compute the class-presence-normalized abnormal/normal weighted mixture. & Makes the final scalar score reproducible once the structured tags are fixed. \\
\bottomrule
\end{tabularx}
\end{table}

\subsection{Stage 1: Report processing}

An \textbf{atomic finding sentence} is a single English clinical statement that expresses exactly one clinical proposition as an (\textbf{anatomical site}, \textbf{finding}) pair. The site identifies \emph{where} the statement applies, and the finding states \emph{what} is observed as either normal or abnormal. RadSEM rewrites all valid Findings-section statements into atomic finding sentences, including both abnormal findings and normal descriptions. Abnormality is not a filter for rewriting; it is introduced later as a class label during matching and as a weight during score aggregation. Report-generation errors are often atomic-sentence-level errors: a generated report may invert a negation, omit a lesion detail, assign a finding to the wrong side, or hallucinate a device or abnormality. Whole-report overlap metrics and unconstrained similarity scores make these local differences difficult to isolate.

\paragraph{Input scope and allowed forms.}
Stage 1 uses only the supplied Findings text. Impressions, history, indications, recommendations, comparison-to-prior wording, technique descriptions, headings, measurements not tied to a finding, and other non-finding content are excluded before scoring. This implements the paper's scoring scope: RadSEM measures agreement in image-observed Findings rather than agreement in context-dependent diagnostic synthesis. The allowed output form is an anatomical location plus one finding, either normal or abnormal. Statements outside this form are removed rather than converted into a looser description.

\paragraph{Splitting rules.}
The rewriting stage splits compound content before filtering. A sentence that contains multiple findings is decomposed until each output sentence contains exactly one finding. A sentence that mixes normal and abnormal content is also split, because normal-background claims and abnormal findings receive different weights downstream. Paired organs are handled with a strong laterality convention: if both sides are mentioned, the sentence is split into left- and right-specific statements; if a paired organ is mentioned without specifying a side, the statement is rewritten into separate left- and right-specific sentences. For MRI reports, sequence-specific observations are isolated so that signal or enhancement descriptions from different sequences do not merge into a single unit.

\paragraph{Filtering and output.}
After splitting, Stage 1 removes content that would not form a scoreable finding. Measurements are retained only when attached to a specific finding. Foreign bodies and medical devices, including tubes, catheters, lines, implants, and similar objects, are treated as abnormal findings when they are identified on imaging. The rewritten sentences are kept in the original report order, completely identical duplicate sentences are removed while retaining the first occurrence, and the result is stored as a machine-readable JSONL record containing the case name, examined area, examined type, and \texttt{rewritten\_report}. Order is not used directly by the scoring formula, but preserving it makes the rewritten report easier to inspect and facilitates manual debugging.

\subsection{Stage 2: Sentence matching with anti-contradiction}

Given $A^{\mathrm{Ref}}$ and $A^{\mathrm{Gen}}$, RadSEM identifies a list of accepted sentence pairs $\mathcal{P}\subseteq A^{\mathrm{Ref}}\times A^{\mathrm{Gen}}$ and unmatched lists $\mathcal{U}^{\mathrm{Ref}}\subseteq A^{\mathrm{Ref}}$ and $\mathcal{U}^{\mathrm{Gen}}\subseteq A^{\mathrm{Gen}}$. The implementation reads two rewritten-report JSONL files, aligns records by case name, and sends the corresponding Ref and Gen rewritten reports to the matching prompt. The alignment is many-to-many: an atomic finding sentence may appear in multiple accepted pairs when the paired statements have substantially the same clinical meaning, including clinically meaningful part--whole or broader--narrower relationships. This policy is intended to absorb residual granularity differences after independent rewriting. For example, an anatomically broad atomic statement may match more specific anatomical components, and a specific statement may match a broader counterpart, when the paired statements refer to the same clinical proposition. Sentences that have no meaningfully similar compatible partner are emitted as unmatched. The pair list is treated as local semantic evidence for the scoring stage, where matched credit is allocated under sentence-level capacity constraints. This is the only stage in which semantic judgment is used for alignment; however, the judgment is constrained to local pair labels rather than an unrestricted overall report score.

\paragraph{Compatibility before similarity.}
The matching decision is based on whether the described entity or object and clinical situation are consistent, or whether one is a clinically meaningful inclusion of the other. Thus, RadSEM does not require exact lexical identity: ``left upper-lobe pulmonary nodule'' may be compatible with a broader ``left lung nodule'' statement, and a general abnormality can match a more specific subtype when the relationship is part--whole. At the same time, semantic relatedness is not sufficient. Compatibility is necessary but not sufficient: an accepted pair must refer to the same finding concept or a clinically meaningful containment relation of that concept, rather than merely describing co-occurring abnormalities in the same anatomy. A pair is accepted only if the two statements could both be true in the same patient at the same time.

\paragraph{Many-to-many pair alignment.}
Many-to-many matching is used to represent semantic coverage across different sentence granularities. A Ref sentence can match multiple Gen sentences, and a Gen sentence can match multiple Ref sentences, when each pair is clinically compatible and expresses the same core proposition at an equivalent or contained granularity. Unsupported additions, independent lesion instances, unrelated findings, and contradictory statements remain unmatched rather than being absorbed by a broad sentence. Stage 3 then scores the emitted pair list deterministically, so flexible alignment is represented through explicit pair-level weights, sentence-level credit capacities, and unmatched-sentence counts rather than through an LLM-assigned global score.

\paragraph{Anti-contradiction constraint.}
A candidate pair $(a_i,b_j)$ is \textbf{forbidden} if both statements cannot be simultaneously true. Typical contradictions include asserted versus negated findings, present versus absent disease, normal versus abnormal status for the same target, opposing directionality such as increased versus decreased, and status opposites such as patent versus obstructed. Forbidden pairs are not counted as matches; the involved sentences remain unmatched unless they find other compatible matches. This rule is a central difference between RadSEM and unconstrained semantic similarity: two sentences can discuss the same anatomy and abnormality while still being clinically incompatible.

\paragraph{Pair-level labels.}
Each accepted pair is represented as a local clinical alignment. The \emph{normality} label is \emph{normal} when the paired sentences describe only normal findings; otherwise it is \emph{abnormal}. The anatomy label records whether the anatomical sites are equivalent or related by part--whole containment. Separate labels compare asserted abnormality concepts and negated abnormality concepts; if neither sentence contains the corresponding asserted or negated abnormality concept, the label is null. This separation helps avoid accidental credit when two sentences mention the same disease term but differ in whether it is present or absent.

\paragraph{Abnormal-detail handling.}
A separate detail label captures information that refines an asserted abnormal finding without changing its core site or concept. Details include measurements, number, severity, morphology, appearance, density, signal, enhancement, acute/chronic descriptors, diagnostic inference phrases, and uncertainty cues. Anatomical localization and the core abnormality term itself are not counted as details. The detail relationship is equivalent, partial, or none, depending on how much abnormal-detail information is shared. If a sentence contains no such descriptors, it is treated as having no details for this comparison. For normal pairs, the abnormal-detail label is null. This design prevents a correct coarse abnormal match from receiving the same credit as a fully specified match, while avoiding brittle exact-string comparison of modifiers.

\paragraph{Unmatched sentences and partial credit.}
Stage 2 also outputs every atomic finding sentence from Ref or Gen that has no meaningfully similar compatible match in the other report. Each unmatched item records whether it comes from Ref or Gen and whether it belongs to the normal or abnormal class. Unmatched Ref sentences become omission-like errors, while unmatched Gen sentences become unsupported-addition or hallucination-like errors. For accepted pairs, part--whole labels provide controlled partial credit when the site or abnormality concept is clinically compatible but differs in granularity. The final score is not produced by the LLM at this stage; it is deferred to the explicit scoring code.

\subsection{Stage 3: Scoring with abnormal-focused weighted F1}

RadSEM assigns each accepted pair a fractional weight $w_p\in(0,1]$, derives class-wise matched credit through a capacity-constrained allocation over accepted pairs, and counts unmatched Ref/Gen atomic finding sentences as errors. Each Ref or Gen atomic sentence provides at most one unit of matched-credit capacity within a class, so repeated links involving the same sentence cannot by themselves create more than one sentence's worth of credit. Unmatched sentences are the explicit unmatched items emitted by the matching stage. The final score places higher weight on abnormal findings. Unlike the first two stages, scoring does not call an LLM; it is a deterministic transformation of the structured tag output.

\paragraph{Part--whole coefficient.}
For each pair, the scoring code counts how many relationship labels indicate part--whole containment. Specifically, it checks the anatomical relationship, the asserted abnormality relationship, and the negated abnormality relationship. Let this count be $k_p$. Each part--whole relation indicates a clinically compatible but less exact correspondence, so RadSEM applies the multiplicative coefficient
\[
 c_{\mathrm{pw}}(p)=(1/3)^{k_p}.
\]
Thus, a pair with equivalent anatomy and equivalent abnormality concepts receives no part--whole penalty, while a pair that is compatible only through one or more containment relations receives progressively less credit.

\paragraph{Detail coefficient.}
The detail label contributes a second coefficient. Null detail labels, including those for normal pairs, receive no detail penalty:
\[
 c_{\mathrm{det}}(p)=\begin{cases}
1.0,&\text{equivalent or null},\\
0.75,&\text{partial},\\
0.5,&\text{none}.
\end{cases}
\]
The pair weight used by the implementation is
\[
 w_p=c_{\mathrm{pw}}(p)c_{\mathrm{det}}(p).
\]
This weighting separates two common sources of partial correctness: a site or abnormality can be clinically related but coarser or finer than the reference, and an abnormality can be correctly detected while still missing important modifiers such as size, severity, morphology, or uncertainty. A broad-to-fine pair therefore receives less credit than an equivalent pair, while unsupported statements remain unmatched. We expose the accepted pairs, labels, and unmatched sentences rather than treating the scalar score as self-explanatory.

\paragraph{Matched and unmatched counts.}
For each class $c\in\{\mathrm{abn},\mathrm{norm}\}$, let $\mathcal{P}_c$ be the accepted pairs labeled with that class. For a pair $p=(a_i,b_j)$, RadSEM allocates a scored contribution $x_p$ bounded by the pair weight and by the available credit capacity of its two sentences:
\[
M_c=\max_{\{x_p\}}\sum_{p\in\mathcal{P}_c}x_p
\]
subject to
\[
0\le x_p\le w_p,\qquad
\sum_{p=(a_i,\cdot)\in\mathcal{P}_c}x_p\le 1\ \forall a_i,\qquad
\sum_{p=(\cdot,b_j)\in\mathcal{P}_c}x_p\le 1\ \forall b_j.
\]
The resulting $M_c$ is the capacity-constrained weighted matched count. The implementation solves this allocation as a fractional bipartite max-flow problem with Ref sentences on one side, Gen sentences on the other side, unit capacities on sentence nodes, and pair weights on Ref--Gen edges. This construction lets many-to-many semantic links remain visible in the tag output while ensuring that a single atomic sentence contributes at most one unit of matched credit per class.

RadSEM also counts raw unmatched sentences in four bins: unmatched Ref abnormal, unmatched Ref normal, unmatched Gen abnormal, and unmatched Gen normal. These raw counts are denoted by $U^{\mathrm{Ref}}_c$ and $U^{\mathrm{Gen}}_c$. Unmatched counts come directly from the unmatched list emitted by Stage 2 rather than from unused candidate-pair credit. Normal bins include ordinary normal observations and negated abnormality statements when they are labeled as normal. This class- and side-specific accounting lets RadSEM penalize both omissions and unsupported generated findings rather than behaving like a recall-only score.

\paragraph{Class-wise F1 and deterministic overrides.}
For each class, the base class-wise score uses the capacity-constrained matched count
\[
\mathrm{F1}^{(0)}_c=
\begin{cases}
\dfrac{2M_c}{2M_c+U^{\mathrm{Ref}}_c+U^{\mathrm{Gen}}_c},&2M_c+U^{\mathrm{Ref}}_c+U^{\mathrm{Gen}}_c>0,\\[0.7em]
0,&\text{otherwise}.
\end{cases}
\]
The implementation then applies a no-unmatched class rule: if $U^{\mathrm{Ref}}_c+U^{\mathrm{Gen}}_c=0$, the intermediate class score is first set to 1.0 during class-wise computation. The final mixture later includes only classes present in the case. RadSEM then applies a no-unmatched partial adjustment for no-unmatched classes that still contain at least one unique matched pair with weight below 1.0. Let $\widetilde{\mathcal{P}}_c$ denote the collection of unique matched Ref--Gen sentence pairs in class $c$, with exact duplicate sentence pairs counted once, and let
\[
\mathcal{W}_c=\bigl[\,w_p: p\in\widetilde{\mathcal{P}}_c\,\bigr]
\]
be the corresponding multiset of pair weights, so repeated weight values from different unique sentence pairs are retained. If $U^{\mathrm{Ref}}_c+U^{\mathrm{Gen}}_c=0$, $\mathcal{W}_c$ is nonempty, and at least one $w\in\mathcal{W}_c$ is below 1.0, RadSEM computes
\[
q_c=\frac{1}{|\mathcal{W}_c|}\sum_{w\in\mathcal{W}_c}w,
\qquad
\lambda_c=\frac{0.25}{\sqrt{|\mathcal{W}_c|}},
\]
and sets
\[
\mathrm{F1}_c=1-\lambda_c(1-q_c),
\]
clipped to the interval $[0,1]$. If the class has any unmatched Ref/Gen sentence, no matched pair, or all unique matched pairs have $w=1.0$, this no-unmatched partial adjustment is not applied. The rule prevents no-unmatched classes containing partial-weight matches from receiving a perfect class score solely because every sentence has a clinically compatible counterpart, while making the adjustment milder as the number of compatible pairs increases.

\paragraph{Final mixture.}
The final case-level score uses abnormal-focused base weights and re-normalizes them over the classes present in the case. Let
\[
\alpha_{\mathrm{abn}}=0.9,\qquad \alpha_{\mathrm{norm}}=0.1,
\]
and let $\mathcal{C}_{\mathrm{present}}$ contain each class that has at least one Ref or Gen atomic finding sentence observed through either an accepted pair or an unmatched item. RadSEM is computed as
\[
\mathrm{RadSEM}=
\begin{cases}
\dfrac{\sum_{c\in\mathcal{C}_{\mathrm{present}}}\alpha_c \mathrm{F1}_c}{\sum_{c\in\mathcal{C}_{\mathrm{present}}}\alpha_c},&\mathcal{C}_{\mathrm{present}}\ne\varnothing,\\[0.9em]
0,&\mathcal{C}_{\mathrm{present}}=\varnothing.
\end{cases}
\]
This mixture reflects the practical assumption that abnormal findings usually dominate clinical utility in report generation, while still retaining a smaller contribution from normal-background statements when normal statements are present. If both abnormal and normal classes are present, the expression reduces to the 0.9/0.1 weighted mixture; if only one class is present, that class receives all of the final weight. The empty-class fallback is a defensive implementation guard for non-scoreable inputs and is not used by the SSREE cases, which contain at least one atomic finding class.

\paragraph{Alternative alignment--allocation policy: RadSEM-Alt.}
To contextualize RadSEM's many-to-many matching design, we also consider a Ref-anchored alternative, denoted \textbf{RadSEM-Alt}. RadSEM-Alt uses the same atomic finding rewriting stage as RadSEM, including duplicate-sentence removal, and keeps the same pair-level relation labels and pair-weight definitions. The difference begins at sentence matching and aggregation. In Stage~2, RadSEM-Alt uses a Ref-anchored one-to-many policy: one Ref sentence may be paired with multiple clinically distinct, non-overlapping, Ref-supported Gen sentences, whereas each Gen sentence may be assigned to at most one best Ref sentence. If a Gen sentence is compatible with multiple Ref candidates, a single best Ref assignment is selected and the remaining Ref sentences are left unmatched unless matched by other Gen sentences. Contradictory statements are not paired.

In Stage~3, RadSEM-Alt defensively enforces this policy by retaining one assignment per Gen sentence, prioritizing higher-weight assignments when conflicts occur, greedily accumulating Gen contributions for each Ref sentence up to a total capacity of one, and truncating the final contribution if it exceeds the remaining Ref capacity. After this filtering step, unmatched Ref and Gen counts are recomputed from the retained assignments. The class-wise F1 calculation, no-unmatched class rule, no-unmatched partial adjustment, and class-presence-normalized abnormal/normal mixture are then applied to the filtered Alt assignments. This variant is used only as a design comparison in Section~\ref{sec:results}; it is not the default RadSEM scoring policy.

\paragraph{Behavior of partial matches and unmatched sentences.}
The class-wise formula makes a useful distinction between reduced matched credit and explicit unmatched-sentence errors. A part--whole coefficient such as $1/3$ does not impose a fixed final-score penalty; it changes the allocated matched credit $M_c$, and the final effect depends on the surrounding amount of matched credit and unmatched sentences. For an illustrative abnormal target finding with no competing accepted links for the target Ref or Gen sentence, let $M_0$ be the existing weighted matched abnormal credit from the remaining findings, excluding the target finding, and let $U_0$ be the existing unmatched abnormal sentence count from the remaining findings. Then, under the base abnormal-class aggregation formula before the no-unmatched override, the no-unmatched partial adjustment, and the final abnormal/normal mixture, the target contributes as follows:
\[
F_{\mathrm{full}}=\frac{2(M_0+1)}{2(M_0+1)+U_0},\qquad
F_{\mathrm{partial}}=\frac{2(M_0+1/3)}{2(M_0+1/3)+U_0},
\]
\[
F_{\mathrm{unmatched},1}=\frac{2M_0}{2M_0+U_0+1},\qquad
F_{\mathrm{unmatched},2}=\frac{2M_0}{2M_0+U_0+2}.
\]
Here $M_0$ is an allocated weighted pair-credit quantity rather than a raw pair or sentence count. In this one-target setting, a full abnormal pair contributes $1$ to $M_0$, whereas a part--whole abnormal pair with one $1/3$ coefficient contributes $1/3$. In contrast, each unmatched Ref or Gen abnormal sentence contributes one raw count to $U_0$.

\begin{figure}[!t]
\centering
\begin{subfigure}[t]{0.49\textwidth}
\centering
\includegraphics[width=\linewidth]{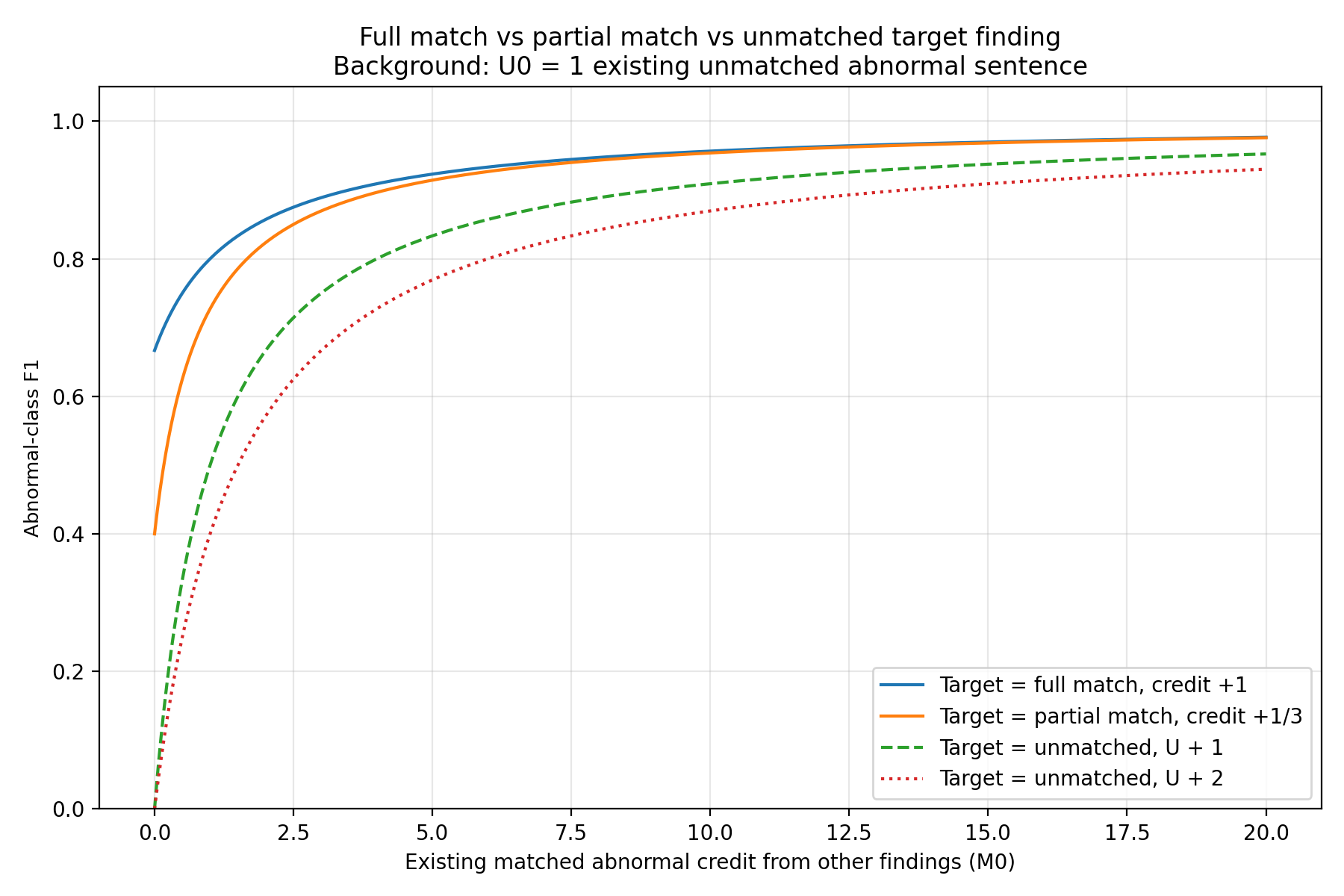}
\caption{F1 trajectories.}
\label{fig:abnormal-f1-scenarios}
\end{subfigure}\hfill
\begin{subfigure}[t]{0.49\textwidth}
\centering
\includegraphics[width=\linewidth]{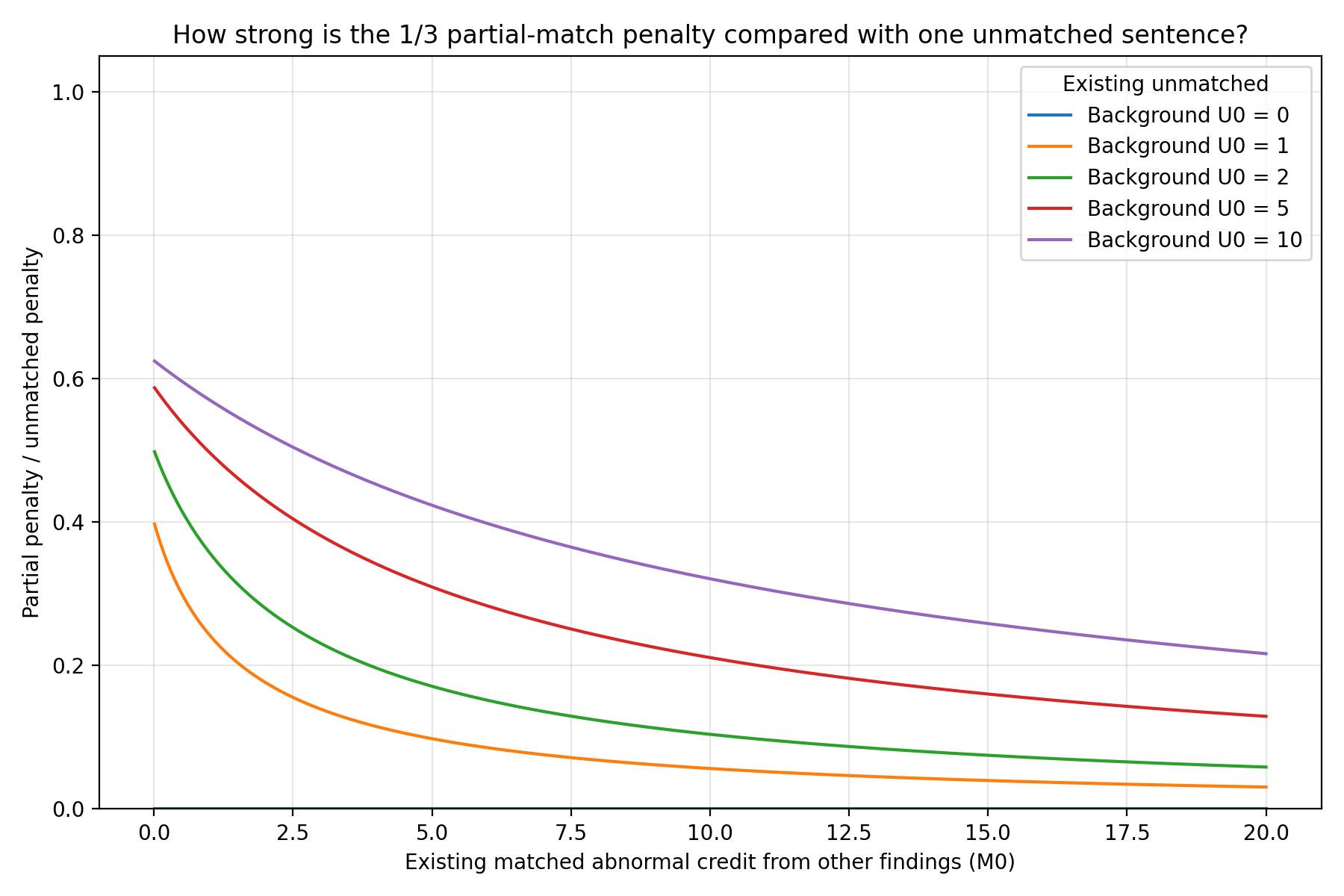}
\caption{Relative penalty ratio.}
\label{fig:partial-penalty-ratio}
\end{subfigure}
\caption{Behavior of partial matches and unmatched sentences under the base abnormal-class aggregation formula. (a) Abnormal-class F1 under full-match, partial-match, and unmatched target-finding scenarios. The background is fixed at $U_0=1$ existing unmatched abnormal sentence, and the horizontal axis varies $M_0$, the existing weighted matched abnormal credit from the remaining findings. The target abnormal finding is then added as a full match with credit $+1$, as a partial match with credit $+1/3$, as one unmatched abnormal sentence, or as two unmatched abnormal sentences. (b) Relative strength of the $1/3$ partial-match penalty compared with making the same target abnormal finding one unmatched abnormal sentence, plotted as $(F_{\mathrm{full}}-F_{\mathrm{partial}})/(F_{\mathrm{full}}-F_{\mathrm{unmatched}})$, where $F_{\mathrm{unmatched}}=2M_0/(2M_0+U_0+1)$. A value of $0.2$ means that assigning a $1/3$ partial-match coefficient causes 20\% as much abnormal-F1 reduction as making the target finding unmatched; a value of $1.0$ would mean an equally severe penalty. The $U_0=0$ curve in panel (b) shows the base formula before the no-unmatched override and no-unmatched partial adjustment, and should not be interpreted as the final class score in no-unmatched cases containing matched pairs with weight below 1.0. Both panels isolate one-target settings in which matched pairs are accumulated as weighted credit in $M$, whereas unmatched findings enter as raw sentence-level errors in $U$; cases with competing many-to-many links are governed by the matched-credit allocation constraints.}
\label{fig:partial-match-behavior}
\end{figure}

Figure~\ref{fig:abnormal-f1-scenarios} and Figure~\ref{fig:partial-penalty-ratio} clarify why RadSEM's partial-credit coefficients should be interpreted as weighted matched credit rather than as direct error penalties. In the one-target setting shown here, a $1/3$ part--whole match generally remains closer to a full match than to an unmatched finding, especially when the report already contains substantial matched abnormal credit. Conversely, unmatched abnormal sentences act as direct denominator terms and can produce larger score changes. In cases with competing links, the allocation constraints determine how much of each compatible pair can contribute to $M_c$. When a class has no unmatched sentences but contains any matched pair with weight below 1.0, the no-unmatched partial adjustment described above uses the average pair weight and the number of unique matched pairs to apply a mild quality discount. This prevents partial granularity or abnormal-detail mismatches from being completely masked by full-credit pairs, while preserving the distinction between partial matches and unmatched errors. Thus, the numerical effect of a partial-match coefficient depends on unmatched context, available sentence-level matched-credit capacity, and the no-unmatched partial adjustment rather than behaving as a fixed final-score decrement.

The no-unmatched partial adjustment makes the scoring policy more internally consistent. A no-unmatched class indicates that all Ref and Gen atomic finding sentences have clinically compatible counterparts, so the class should not be penalized as strongly as one containing omissions or unsupported additions. However, if any accepted pair has a weight below 1.0, the class still contains a granularity or abnormal-detail mismatch. RadSEM therefore applies a mild adjustment based on the average unique-pair weight, preventing partial matches from being completely masked by full-credit pairs while preserving the distinction between partial matches and unmatched errors.

\FloatBarrier

\paragraph{Reproducibility and auditability.}
Stages 1--2 use constrained structured outputs containing atomic finding sentences, accepted pairs, pair labels, and unmatched sets; Stage 3 is fully deterministic given those outputs. RadSEM therefore still depends on LLM rewriting and matching, but changes the model's role from free-form judge to controlled parser and semantic matcher. Users can inspect the rewritten reports, accepted pairs, pair labels, unmatched sentences, raw pair weights, matched-credit allocation, class-wise matched counts, no-unmatched class rule and no-unmatched partial adjustment cases to determine whether a low score was caused by omitted findings, unsupported generated findings, contradictory polarity, anatomy granularity, abnormal-concept granularity, or missing abnormal details.

\section{Experiments}

\subsection{Dataset and setting}

We randomly sampled 3,000 real radiology reports from an internal tertiary-hospital repository restricted to the most recent three years. Sampling was stratified by modality, including MRI, computed tomography (CT), and computed radiography (CR), to match the in-database modality distribution. To avoid degenerate stress-test cases, we retained reports that contained at least three abnormal and at least one normal atomic finding sentence after standardization. This constraint ensures that all five SSREE levels can be constructed without collapsing into one another; for example, if a report had only two abnormal sentences, the L4 rewrite, in which all abnormal sentences are converted to normal statements, would be indistinguishable from L3. After filtering, 2,448 reports remained for the experiments.

For each original report, we constructed five controlled variants (L1--L5) with increasing semantic deviation and scored each variant against the original reference. This yields $2{,}448\times 5=12{,}240$ reference--candidate scoring instances. For adjacent-level tests, each report contributes four ordered comparisons $(L1,L2)$, $(L2,L3)$, $(L3,L4)$, and $(L4,L5)$, totaling $2{,}448\times 4=9{,}792$ adjacent pairs.

\paragraph{Data source and privacy.}
All samples were derived from real clinical radiology reports. Before use, the report text was thoroughly de-identified and translated into English. Direct identifiers and patient-specific metadata, including patient names or IDs, exact exam dates, referrer or reader names, and other information that could reveal personal identity, were removed. The dataset used in this study contains de-identified report text only and does not include medical images or identifiable patient information.

\subsection{SSREE: monotonicity stress test}

SSREE (Sentence-Sensitive Report Evaluation Experiment) evaluates whether a metric responds monotonically to controlled increases in semantic error severity. It is intended as a controlled semantic stress test, not as a substitute for external clinical validation on real model outputs. Because SSREE edits reports at the finding-unit level, it is naturally aligned with the type of error that RadSEM is designed to measure; we therefore interpret it as evidence about monotonic sensitivity under controlled perturbations rather than as a complete clinical benchmark. A metric passes the stress test when it produces
\[
\mathrm{Score}(L_1)>\mathrm{Score}(L_2)>\mathrm{Score}(L_3)>\mathrm{Score}(L_4)>\mathrm{Score}(L_5)
\]
when each level is compared against the same original reference. To reduce lexical artifacts, SSREE applies randomized synonym substitutions at every level. For L1, these substitutions are intended to preserve clinical meaning while changing wording and organization; for L2--L5, they are combined with the level-specific semantic edits so that performance reflects clinical meaning changes rather than superficial wording.

\begin{figure}[!htbp]
\centering
\includegraphics[width=0.78\textwidth]{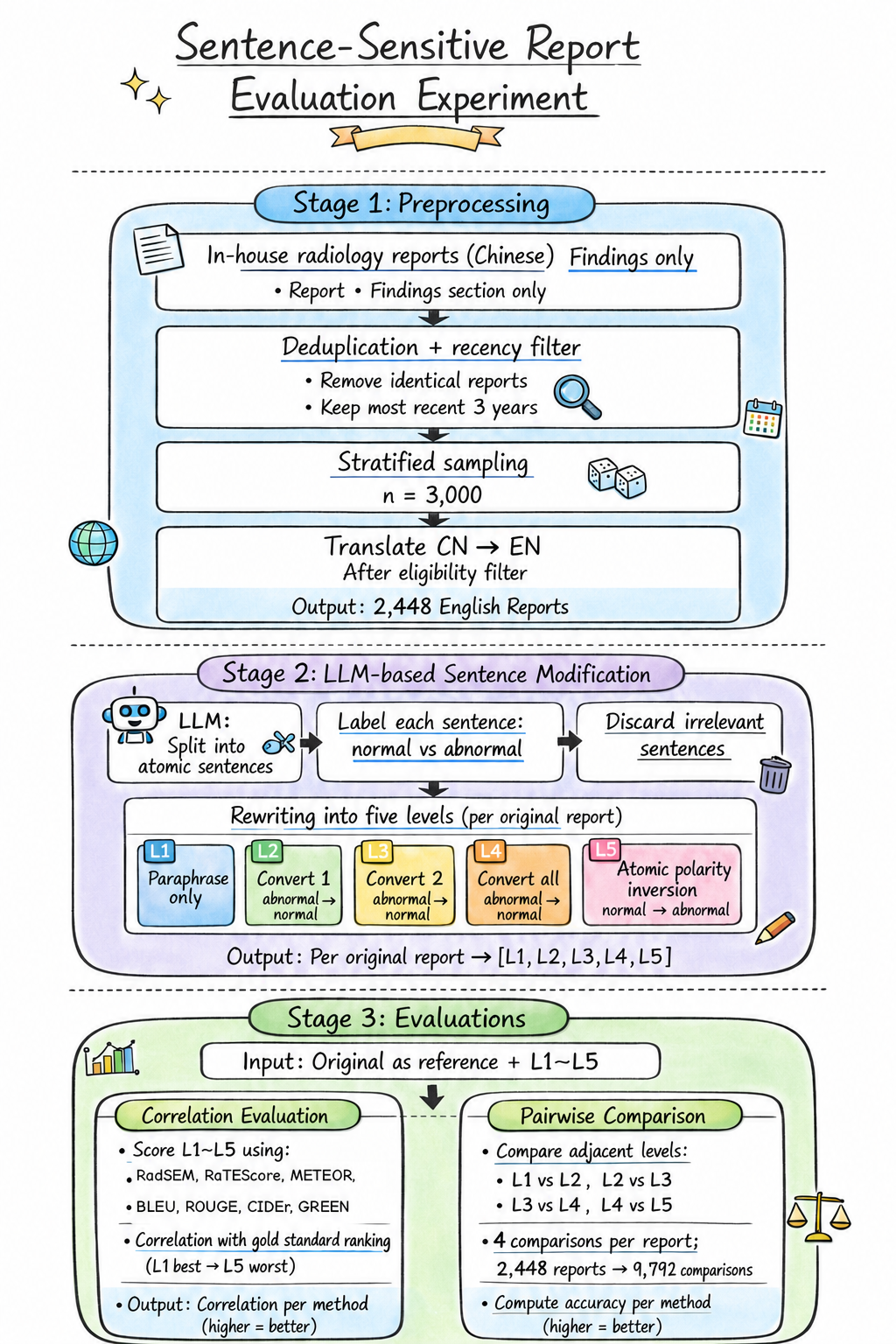}
\caption{SSREE workflow. In Stage 1, modality-stratified sampling produced an initial set of 3,000 recent in-house radiology Findings reports, and atomic-finding eligibility filtering retained 2,448 English reports for SSREE construction. In Stage 2, reports were split into atomic sentences, labeled as normal or abnormal, and rewritten into five controlled levels; L5 applies atomic polarity inversion between normal and abnormal findings. In Stage 3, the original report is used as the reference and L1--L5 are evaluated by correlation with the designed severity order and by adjacent-level pairwise comparisons.}
\label{fig:ssree-workflow}
\end{figure}

\subsubsection{Report standardization}
Before the levels are generated, each original report is standardized once. First, the report is converted into atomic finding sentences under the constraints described in RadSEM Stage~1 (report processing): one finding per sentence, left/right separation for paired organs, sequence isolation when needed, removal of non-finding content, and treatment of foreign bodies and devices as abnormal findings. Level-specific SSREE perturbations are then applied to this atomic sentence set rather than directly to the original paragraph. After the controlled edits are made, the edited sentence set is recombined by GPT-5 into a coherent, natural report-style Findings text for scoring. Recombination is treated as a presentation step: it is instructed to preserve the edited atomic clinical content while improving report-level fluency, rather than introducing an additional source of semantic corruption. The same standardized sentence set is reused to generate all L1--L5 variants.

\subsubsection{Level construction}
Table~\ref{tab:ssree-levels} summarizes the five corruption levels. L1 preserves clinical meaning while varying wording, formatting, and sentence organization. L2--L4 progressively remove abnormal content by converting abnormal statements into corresponding normal statements with the same intended clinical target when possible. L5 performs strict sentence-by-sentence polarity inversion and therefore represents the most severe semantic corruption.

\begin{table}[H]
\caption{SSREE corruption levels with increasing semantic deviation.}
\label{tab:ssree-levels}
\centering
{\small
\begin{tabular}{@{}lp{0.78\textwidth}@{}}
\toprule
Level & Description \\
\midrule
L1 & Apply meaning-preserving synonym substitutions, reordering, and formatting into a coherent report. \\
L2 & Convert exactly one abnormal atomic finding sentence to a corresponding normal statement, preserving the intended clinical target when possible and removing abnormal-only details incompatible with normality. \\
L3 & Same as L2 but convert two distinct abnormal atomic finding sentences. \\
L4 & Convert all abnormal sentences to normal, including devices and foreign bodies. \\
L5 & Strict inversion for every sentence: normal $\leftrightarrow$ abnormal, preserving sentence count, boundaries, and clinically relevant details. \\
\bottomrule
\end{tabular}}
\end{table}

This design yields a controlled severity ladder: L1 changes wording and organization without changing clinical meaning; L2--L4 progressively remove abnormal content; and L5 inverts polarity throughout the report. Because L2--L4 are generated as natural report text after atomic-level editing, the corresponding normal statements may use conventional anatomical wording appropriate for report-style expression, while the intended clinical target remains aligned with the edited atomic finding. Internal checks enforce that L5 maintains the original sentence count and that clinically relevant details are inverted rather than dropped.

In abnormal-to-normal polarity flips, RadSEM records the missed reference abnormality in the abnormal class and the unsupported generated normal statement in the normal class. This accounting is intentional rather than an unintended double penalty: when both classes are present, the clinically primary penalty is carried by the abnormal class through the 0.9 base mixture weight, while the normal-side penalty is a low-weight consequence of explicitly exposing the generated statement that contradicts the reference abnormality.

\paragraph{Qualitative example.}
Table~\ref{tab:l3-case-complete} shows a representative L3 case from a chest radiograph report. The L3 rewrite preserves the thoracic cage, lung-field transparency, bilateral linear and patchy shadows, pulmonary markings, hilar density, cardiac silhouette, diaphragmatic surfaces, and costophrenic angles, while converting two abnormal findings into corresponding normal statements: mild scoliosis of the spine and prominence of the aortic knob. This case provides a compact but clinically interpretable audit trail because the two injected polarity errors are isolated among many correctly matched normal and abnormal findings.

{\fontsize{5.85pt}{6.45pt}\selectfont
\setlength{\LTpre}{1.15em}
\setlength{\LTpost}{0.25em}
\setlength{\tabcolsep}{2.6pt}
\renewcommand{\arraystretch}{0.92}
\begin{longtable}{@{}>{\raggedright\arraybackslash}p{0.47\textwidth}>{\raggedright\arraybackslash}p{0.47\textwidth}@{}}
\caption{Qualitative L3 case showing RadSEM error localization for a chest radiograph report. The first block shows the complete text used for the example; the matched and unmatched atomic finding sentences below are reproduced verbatim from the tag output. $M$ denotes the capacity-constrained weighted matched count, \textit{abn} denotes abnormal, and \textit{norm} denotes normal; in this example, the matched links use their full displayed credit and the error counts equal the raw unmatched counts.}\label{tab:l3-case-complete}\\
\toprule
\textbf{Generated L3 report} & \textbf{Reference report} \\
\midrule
\multicolumn{2}{@{}l}{\textbf{Complete report text}}\\[0.12em]
\cmidrule(lr){1-2}
Thoracic cage is symmetric. Left lung field is normally lucent. Right lung field is normally lucent. Left lung field contains linear opacities. Right lung field contains linear opacities. Left lung field contains patchy opacities. Right lung field contains patchy opacities. Left lung pulmonary markings are well delineated. Right lung pulmonary markings are well delineated. Left pulmonary hilum is not enlarged. Right pulmonary hilum is not enlarged. Left pulmonary hilum shows no increased density. Right pulmonary hilum shows no increased density. Cardiac silhouette has normal configuration. Cardiac silhouette is normal in size. Left hemidiaphragmatic contour is smooth. Right hemidiaphragmatic contour is smooth. Left costophrenic angle is sharply demarcated. Right costophrenic angle is sharply demarcated. Spine shows no scoliotic curvature. Aortic knob shows no prominence.
&
The thoracic cage is symmetrical on both sides. The transparency of both lung fields is normal. Striated and patchy shadows are seen in both lung fields. Pulmonary markings are clear. No enlargement or increased density is observed in both pulmonary hila. The cardiac silhouette is normal in shape and size. Both diaphragmatic surfaces are smooth, and the costophrenic angles are sharp. Mild scoliosis of the spine is noted. The aortic knob is prominent.
\\
\midrule
\addlinespace[0.18em]
\multicolumn{2}{@{}l}{\textbf{Atomic finding sentence list used by RadSEM}}\\[0.12em]
\cmidrule(lr){1-2}
\textbf{Generated L3 atomic finding sentence} & \textbf{Reference atomic finding sentence} \\
\midrule
Thoracic cage is symmetric. \newline
Left lung field is normally lucent. \newline
Right lung field is normally lucent. \newline
Left lung field contains linear opacities. \newline
Right lung field contains linear opacities. \newline
Left lung field contains patchy opacities. \newline
Right lung field contains patchy opacities. \newline
Left lung pulmonary markings are well delineated. \newline
Right lung pulmonary markings are well delineated. \newline
Left pulmonary hilum is not enlarged. \newline
Right pulmonary hilum is not enlarged. \newline
Left pulmonary hilum shows no increased density. \newline
Right pulmonary hilum shows no increased density. \newline
Cardiac silhouette has normal configuration. \newline
Cardiac silhouette is normal in size. \newline
Left hemidiaphragmatic contour is smooth. \newline
Right hemidiaphragmatic contour is smooth. \newline
Left costophrenic angle is sharply demarcated. \newline
Right costophrenic angle is sharply demarcated. \newline
Spine shows no scoliotic curvature. \newline
Aortic knob shows no prominence.
&
Thoracic cage is symmetrical. \newline
Left lung field transparency is normal. \newline
Right lung field transparency is normal. \newline
Left lung field has striated shadows. \newline
Right lung field has striated shadows. \newline
Left lung field has patchy shadows. \newline
Right lung field has patchy shadows. \newline
Left lung field pulmonary markings are clear. \newline
Right lung field pulmonary markings are clear. \newline
Left pulmonary hilum is not enlarged. \newline
Right pulmonary hilum is not enlarged. \newline
Left pulmonary hilum does not have increased density. \newline
Right pulmonary hilum does not have increased density. \newline
Cardiac silhouette shape is normal. \newline
Cardiac silhouette size is normal. \newline
Left diaphragmatic surface is smooth. \newline
Right diaphragmatic surface is smooth. \newline
Left costophrenic angle is sharp. \newline
Right costophrenic angle is sharp. \newline
Spine has mild scoliosis. \newline
Aortic knob is prominent.
\\
\midrule
\addlinespace[0.18em]
\multicolumn{2}{@{}l}{\textbf{Matched atomic finding sentence pairs}}\\[0.12em]
\cmidrule(lr){1-2}
\textbf{Generated atomic finding sentence} & \textbf{Reference atomic finding sentence} \\
\midrule
Thoracic cage is symmetric. & Thoracic cage is symmetrical. \\
Left lung field is normally lucent. & Left lung field transparency is normal. \\
Right lung field is normally lucent. & Right lung field transparency is normal. \\
Left lung field contains linear opacities. & Left lung field has striated shadows. \\
Right lung field contains linear opacities. & Right lung field has striated shadows. \\
Left lung field contains patchy opacities. & Left lung field has patchy shadows. \\
Right lung field contains patchy opacities. & Right lung field has patchy shadows. \\
Left lung pulmonary markings are well delineated. & Left lung field pulmonary markings are clear. \\
Right lung pulmonary markings are well delineated. & Right lung field pulmonary markings are clear. \\
Left pulmonary hilum is not enlarged. & Left pulmonary hilum is not enlarged. \\
Right pulmonary hilum is not enlarged. & Right pulmonary hilum is not enlarged. \\
Left pulmonary hilum shows no increased density. & Left pulmonary hilum does not have increased density. \\
Right pulmonary hilum shows no increased density. & Right pulmonary hilum does not have increased density. \\
Cardiac silhouette has normal configuration. & Cardiac silhouette shape is normal. \\
Cardiac silhouette is normal in size. & Cardiac silhouette size is normal. \\
Left hemidiaphragmatic contour is smooth. & Left diaphragmatic surface is smooth. \\
Right hemidiaphragmatic contour is smooth. & Right diaphragmatic surface is smooth. \\
Left costophrenic angle is sharply demarcated. & Left costophrenic angle is sharp. \\
Right costophrenic angle is sharply demarcated. & Right costophrenic angle is sharp. \\
\midrule
\addlinespace[0.18em]
\multicolumn{2}{@{}l}{\textbf{Unmatched atomic finding sentences}}\\[0.12em]
\cmidrule(lr){1-2}
\textbf{Generated unmatched normal sentence} & \textbf{Reference unmatched abnormal sentence} \\
\midrule
Spine shows no scoliotic curvature. & Spine has mild scoliosis. \\
Aortic knob shows no prominence. & Aortic knob is prominent. \\
\midrule
\addlinespace[0.18em]
\multicolumn{2}{@{}l}{\textbf{Case score}}\\[0.12em]
\cmidrule(lr){1-2}
\multicolumn{2}{@{}p{0.94\textwidth}@{}}{The tag output contains $M_{\mathrm{abn}}=4$, $U^{\mathrm{Ref}}_{\mathrm{abn}}=2$, and $U^{\mathrm{Gen}}_{\mathrm{abn}}=0$, giving $\mathrm{F1}_{\mathrm{abn}}=8/(8+2)=0.800$. It also contains $M_{\mathrm{norm}}=15$, $U^{\mathrm{Ref}}_{\mathrm{norm}}=0$, and $U^{\mathrm{Gen}}_{\mathrm{norm}}=2$, giving $\mathrm{F1}_{\mathrm{norm}}=30/(30+2)=0.938$. Because both classes are present, RadSEM $=0.9\times0.800+0.1\times0.938=\textbf{0.814}$.}\\
\bottomrule
\end{longtable}
}

This tag-level case shows how RadSEM separates clinically compatible paraphrases from contradicted normalizations. Bilateral lung opacities, pulmonary markings, hilar findings, cardiac silhouette, diaphragmatic surfaces, and costophrenic angles are accepted as matched findings, while the generated denials of spinal scoliosis and aortic-knob prominence are not allowed to match the corresponding abnormal reference findings.

\FloatBarrier
\subsection{RaTE-Eval synthetic rewrite comparison}

We also reuse the RaTE-Eval ``Comparison of Synthetic Reports'' subset from Zhao et al.~\citep{zhao2024ratescore}, which originally contains 847 triplets: original, synonymous rewrite, and antonymous rewrite. During quality control, we exclude invalid samples, including failed rewrites that do not preserve the intended synonymous or antonymous relation and instances whose content is not part of the imaging report Findings, such as image quality-control text or patient baseline descriptions. This yields 599 triplets that describe imaging findings and are suitable for scoring.

A metric is counted as correct if it assigns a strictly higher similarity score to the synonymous rewrite than to the antonymous rewrite when both are compared against the original sentence; ties are counted as failures.

\subsection{Baselines and assessment criteria}

We compare RadSEM against RaTEScore, METEOR, BLEU-1/2/3/4, ROUGE-1/2/L, CIDEr, and GREEN. All metrics use the same reference--candidate pairing, scoring each $L_k$ variant against the original report.

SSREE evaluates both local sensitivity and global consistency. For \textbf{adjacent-level accuracy}, each adjacent pair $(L_k,L_{k+1})$ is correct if $\mathrm{Score}(L_k)>\mathrm{Score}(L_{k+1})$; ties and reversals are failures. For \textbf{perfect monotonicity}, each report must satisfy the full strict chain $\mathrm{Score}(L_1)>\mathrm{Score}(L_2)>\mathrm{Score}(L_3)>\mathrm{Score}(L_4)>\mathrm{Score}(L_5)$. For \textbf{Kendall's $\tau_b$}, we compute within-report rank correlation between the expected quality order and metric scores using all ten level pairs with tie correction, and report the sign-adjusted value so that 1.0 indicates perfect agreement with the SSREE quality order. For \textbf{all-pairs concordance}, all ten pairs $(k,\ell)$ with $k<\ell$ are counted as concordant if $s_{i,k}>s_{i,\ell}$, discordant if $s_{i,k}<s_{i,\ell}$, and tied otherwise; ties receive half credit:
\[
\mathrm{Concordance}_i=\frac{C+0.5T}{10}.
\]
Aggregates are computed over reports. Kendall's $\tau_b$ is reported as a correlation coefficient, whereas concordance and accuracy metrics are reported as percentages in the Results section; differences between two percentage-based metrics are reported in percentage points (pp). When confidence intervals are needed, we use nonparametric bootstrap resampling over reports.

\paragraph{RadSEM implementation details.}
Unless specified otherwise, RadSEM uses the class-presence-normalized F1 mixture defined in Section~\ref{sec:method}, with base class weights 0.9 for abnormal findings and 0.1 for normal findings. We use the part--whole penalty $c_{\mathrm{pw}}(p)=(1/3)^{k_p}$, abnormal-detail weights of 1.0, 0.75, and 0.5 for equivalent, partial, and none, capacity-constrained matched-credit allocation, the no-unmatched class rule, and the no-unmatched partial adjustment rule described in Section~\ref{sec:method}. The no-unmatched partial adjustment is triggered whenever a class has no unmatched Ref/Gen sentences and at least one unique matched pair with weight below 1.0. In the released pipeline, Stage~1 rewriting and Stage~2 matching used GPT-5 with fixed prompts and a maximum generation length of 16,392 tokens; SSREE report-style recombination was also performed with GPT-5. Devices and foreign bodies are treated as abnormal during both rewriting and level generation.

\paragraph{RadSEM-Alt comparison.}
We additionally compare RadSEM with RadSEM-Alt as an alternative alignment--allocation policy rather than as an external baseline. This comparison uses the same 2,448 SSREE report groups as the main SSREE evaluation, with valid scores for all five L1--L5 levels for both methods. Thus, the RadSEM versus RadSEM-Alt comparison is performed on the identical sample set used for the external-baseline SSREE evaluation, rather than on a smaller subset.

\section{Results}
\label{sec:results}

\begin{table}[H]
\caption{SSREE ranking consistency over 2,448 reference reports, each expanded into a five-level set (L1--L5) and evaluated via within-report ranking. Kendall $\tau_b$ is reported as a correlation coefficient; all concordance and accuracy measures are reported as percentages. Higher is better.}
\label{tab:ssree-summary}
\centering
{\small
\begin{tabular}{lcccc}
\toprule
Metric & Kendall $\tau_b$ & All-pairs (\%) & Adjacent (\%) & Perfect chain (\%) \\
\midrule
\textbf{RadSEM (ours)} & \textbf{0.957} & \textbf{97.8} & \textbf{95.0} & \textbf{81.9} \\
RaTEScore & 0.597 & 79.8 & 70.5 & 22.4 \\
BLEU-1 & 0.555 & 77.0 & 63.4 & 7.7 \\
BLEU-2 & 0.551 & 77.2 & 64.2 & 10.7 \\
GREEN & 0.525 & 66.0 & 55.7 & 5.3 \\
BLEU-3 & 0.482 & 73.8 & 62.8 & 9.0 \\
ROUGE-1 & 0.448 & 69.3 & 54.7 & 3.9 \\
ROUGE-2 & 0.427 & 70.5 & 60.7 & 6.7 \\
METEOR & 0.422 & 70.9 & 59.9 & 5.8 \\
BLEU-4 & 0.419 & 70.7 & 60.9 & 8.0 \\
ROUGE-L & 0.354 & 65.1 & 52.2 & 2.5 \\
CIDEr & 0.267 & 61.3 & 55.6 & 4.9 \\
\bottomrule
\end{tabular}}
\end{table}

\begin{figure}[!t]
\centering
\includegraphics[width=0.95\textwidth]{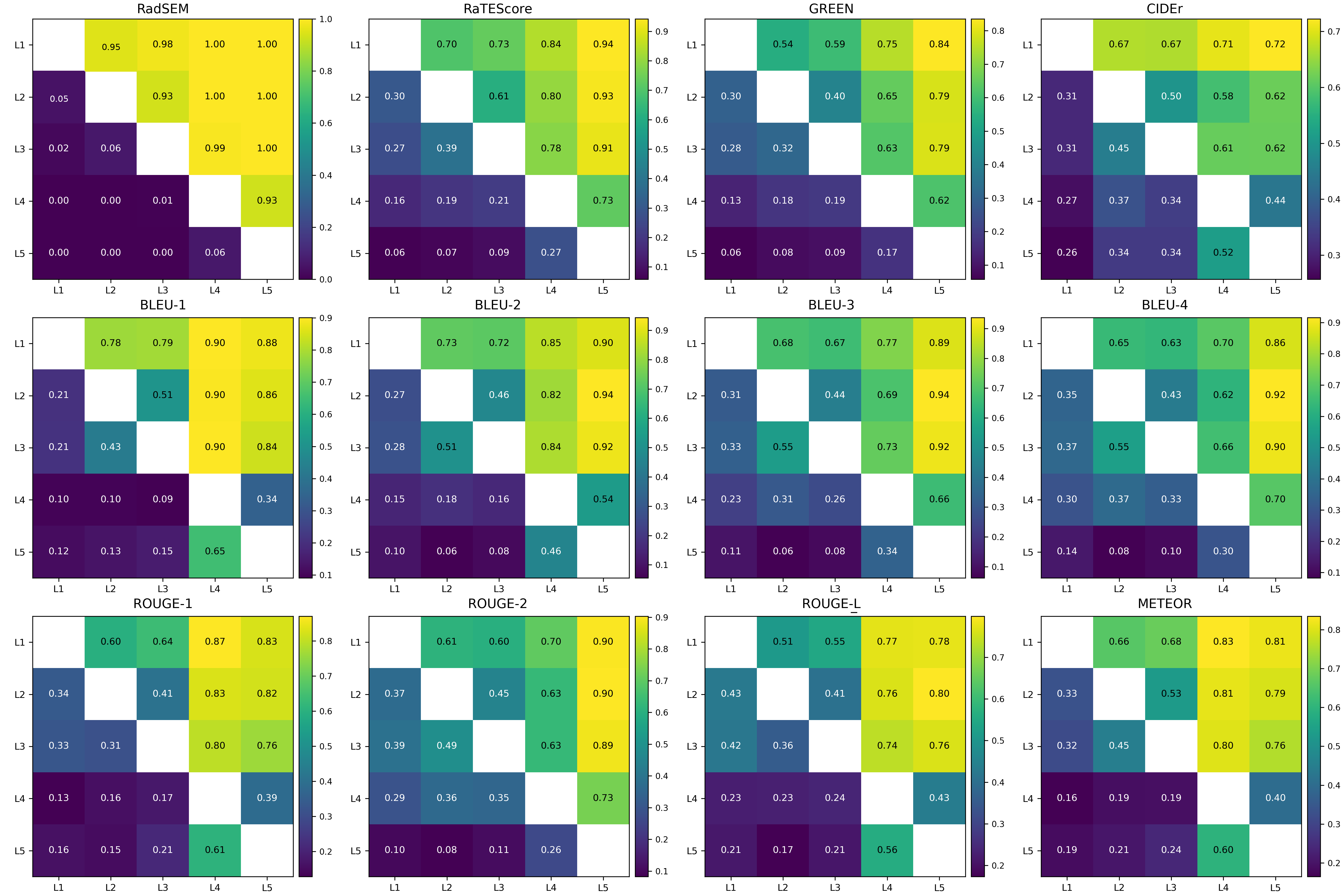}
\caption{SSREE pairwise ordering heatmaps. Each cell reports the fraction of reports for which the metric score of the row level ($Y$) is strictly higher than the score of the column level ($X$), i.e., $P(\mathrm{score}(Y)>\mathrm{score}(X))$. Because SSREE defines report quality as $L1>L2>L3>L4>L5$, an ideal metric forms a strong upper-triangular pattern.}
\label{fig:ssree-heatmaps}
\end{figure}

RadSEM remains consistently monotone under graded corruption, achieving high adjacent discrimination and all-pairs concordance (Table~\ref{tab:ssree-summary}, Fig.~\ref{fig:ssree-heatmaps}). Overlap-based metrics can violate the L4--L5 boundary because L5 reintroduces abnormal tokens via inversion and may increase surface overlap despite being clinically contradictory. RadSEM avoids this failure by forbidding contradictory matches and emphasizing abnormal findings.

\paragraph{RadSEM shows high monotonic consistency.}
RadSEM produces a strongly triangular pattern in the heatmaps: for quality-consistent pairs $(L_i,L_j)$ with $i<j$, the win rates are consistently high, typically in the range of 93\%--100\%, while reverse comparisons remain near zero, typically below 7\%. In particular, RadSEM maintains high adjacent discrimination with $P(L1>L2)=95.1\%$, $P(L2>L3)=93.3\%$, $P(L3>L4)=98.9\%$, and $P(L4>L5)=92.6\%$, yielding an adjacent-pair concordance of 95.0\%. Across all ten level pairs, RadSEM reaches an all-pairs concordance of 97.8\% and Kendall rank consistency of 0.957. It satisfies the strict five-level ordering for 81.9\% of reports, indicating that most reports show a fully monotone score chain rather than only local separability.

\begin{table}[!t]
\caption{Paired comparison between RadSEM and RadSEM-Alt on the full SSREE set of 2,448 reference reports. Both methods are evaluated on the same report groups and the same L1--L5 instances. Overall metrics summarize within-report ranking over the five quality levels; transition metrics report strict adjacent-level ordering accuracy, $P(\mathrm{score}(L_i)>\mathrm{score}(L_{i+1}))$. Kendall $\tau_b$ is reported as a correlation coefficient; all accuracy and concordance measures are reported as percentages, and the $\Delta$ row reports percentage-point (pp) gains for those measures. Higher is better.}
\label{tab:radsem-alt-comparison}
\centering
\begin{threeparttable}
{\small
\resizebox{\textwidth}{!}{%
\begin{tabular}{lcccccccc}
\toprule
& \multicolumn{4}{c}{Overall SSREE ranking} & \multicolumn{4}{c}{Adjacent transition accuracy} \\
\cmidrule(lr){2-5} \cmidrule(lr){6-9}
Method & Kendall $\tau_b$ & All-pairs (\%) & Mean adjacent (\%) & Perfect chain (\%) & L1$>$L2 (\%) & L2$>$L3 (\%) & L3$>$L4 (\%) & L4$>$L5 (\%) \\
\midrule
RadSEM-Alt & 0.935 & 95.8 & 90.7 & 67.2 & 91.5 & 88.2 & 98.8 & 84.1 \\
\textbf{RadSEM} & \textbf{0.957} & \textbf{97.8} & \textbf{95.0} & \textbf{81.9} & \textbf{95.1} & \textbf{93.3} & \textbf{98.9} & \textbf{92.6} \\
\midrule
$\Delta$ RadSEM--Alt & +0.022 & +2.0 pp & +4.3 pp & +14.7 pp & +3.6 pp & +5.1 pp & +0.1 pp & +8.5 pp \\
\bottomrule
\end{tabular}}}
\begin{tablenotes}[flushleft]
\footnotesize
\item \parbox[t]{0.96\textwidth}{Note: The $\Delta$ row reports absolute improvements. For percentage-based columns, improvements are percentage points (pp), not relative percent changes. ``Mean adjacent'' is the average strict ordering accuracy across the four adjacent transitions shown on the right. Percentages and pp values are rounded to one decimal place. RadSEM-Alt is evaluated as a Ref-anchored alternative alignment--allocation policy that shares Stage~1 with RadSEM but uses its own Stage~2 matching and Stage~3 aggregation rules.}
\end{tablenotes}
\end{threeparttable}
\end{table}

\paragraph{Comparison with the Ref-anchored alternative.}
RadSEM also outperforms RadSEM-Alt on the full 2,448-report SSREE set, using the same samples as the main SSREE evaluation (Table~\ref{tab:radsem-alt-comparison}). The paired layout in Table~\ref{tab:radsem-alt-comparison} reports one row per method so that the overall ranking metrics and their adjacent-transition components share a single denominator and can be compared directly. RadSEM improves Kendall $\tau_b$ from 0.935 to 0.957, all-pairs concordance from 95.8\% to 97.8\% (+2.0 pp), mean adjacent concordance from 90.7\% to 95.0\% (+4.3 pp), and strict five-level ordering from 67.2\% to 81.9\% (+14.7 pp). Among the four adjacent transitions, the largest absolute gain appears at the L4--L5 boundary, where RadSEM improves ordering accuracy from 84.1\% to 92.6\% (+8.5 pp).

This comparison should be interpreted as a design-rationale analysis rather than as a causal ablation of a single isolated component, because RadSEM and RadSEM-Alt differ in several coupled alignment and aggregation choices. The results nevertheless support the reasonableness of RadSEM's many-to-many candidate-link formulation. In radiology reports, a generated sentence may partially correspond to multiple more specific reference findings, and several generated fragments may jointly describe one reference finding. RadSEM preserves such compatible links as candidates and resolves the resulting competition during deterministic scoring. RadSEM-Alt instead requires earlier single-best Gen-to-Ref decisions and then applies local filtering with unmatched-count recomputation. This more restrictive policy can convert granularity ambiguity or local matching noise into unmatched penalties, which is consistent with its weaker perfect-order rate and reduced L4--L5 separation.

\paragraph{RaTEScore is the strongest baseline but degrades on fine-grained steps.}
RaTEScore shows reasonable separation for large gaps, for example $P(L1>L5)=94\%$, but weaker sensitivity on adjacent levels, especially early steps. Its $P(L1>L2)=70\%$ and $P(L2>L3)=61\%$ lead to an adjacent-pair concordance of 70.5\% and a strict perfect-order rate of 22.4\%. The heatmap also shows non-trivial reversals, such as $P(L5>L4)=27\%$, suggesting instability when the perturbation changes the polarity and distribution of abnormal statements.

\paragraph{Lexical overlap metrics exhibit systematic L4--L5 failures.}
Across BLEU, ROUGE, METEOR, and CIDEr, the upper triangle is often moderately high for comparisons involving L1, but a recurring failure mode appears at the L4--L5 boundary. Several overlap-based metrics assign higher scores to L5 than to L4 in a large fraction of cases: BLEU-1 yields $P(L5>L4)=65\%$, ROUGE-1 yields 61\%, ROUGE-L yields 56\%, METEOR yields 60\%, and CIDEr yields 52\%. This behavior is consistent with SSREE construction: L4 normalizes abnormalities away, reducing overlap with an often abnormal reference, while L5 reintroduces abnormal tokens through inversion, increasing surface overlap despite being clinically contradictory. These metrics therefore struggle to enforce the intended severity ladder, producing low strict perfect-order rates, typically 2.5\%--10.7\%, and modest adjacent concordance, roughly 52\%--64\%.

\paragraph{GREEN and CIDEr are unstable on subtle perturbations.}
GREEN and CIDEr show particularly weak discrimination on subtle perturbations. GREEN's $P(L1>L2)=54\%$ is close to chance, and CIDEr's adjacent performance is similarly limited, with 55.6\% adjacent concordance and 4.9\% strict perfect order. Their heatmaps contain both missed separations in the upper triangle and substantial mass in the lower triangle, indicating frequent violations of the expected ranking.

\begin{table}[!t]
\caption{Accuracy on the RaTE-Eval synthetic rewrite comparison subset (599 triplets). A trial is correct if $\mathrm{Score}(\text{original},\text{synonymous})>\mathrm{Score}(\text{original},\text{antonymous})$.}
\label{tab:rateeval-synthetic}
\centering
{\small
\begin{tabular}{lrrr}
\toprule
Metric & Total & Correct & Accuracy \\
\midrule
\textbf{RadSEM (ours)} & 599 & 597 & 99.67\% \\
GREEN & 599 & 587 & 98.00\% \\
RaTEScore & 599 & 374 & 62.44\% \\
BLEU-1 & 599 & 338 & 56.43\% \\
METEOR & 599 & 314 & 52.42\% \\
BLEU-2 & 599 & 307 & 51.25\% \\
BLEU-3 & 599 & 295 & 49.25\% \\
BLEU-4 & 599 & 293 & 48.91\% \\
ROUGE-L & 599 & 283 & 47.25\% \\
CIDEr & 599 & 278 & 46.41\% \\
ROUGE-1 & 599 & 268 & 44.74\% \\
ROUGE-2 & 599 & 249 & 41.57\% \\
\bottomrule
\end{tabular}}
\end{table}

\paragraph{Synthetic rewrite comparison.}
On the RaTE-Eval synthetic rewrite subset, RadSEM correctly prefers synonymous rewrites over antonymous rewrites for 597 of 599 triplets, achieving 99.67\% accuracy (Table~\ref{tab:rateeval-synthetic}). GREEN is also strong on this binary comparison at 98.00\%, while RaTEScore and generic text metrics are lower. This result complements SSREE by showing that the same matching mechanism is also sensitive to sentence-level synonym/antonym perturbations.

\paragraph{Summary.}
Overall, SSREE shows that RadSEM is more consistent with the desired monotonic response to graded semantic corruption than the evaluated baselines. It shows strong adjacent sensitivity and long-range ordering, and avoids the common L4--L5 inversion failure of overlap-driven metrics. This supports the intended design goal of being sensitive to clinically dangerous polarity flips and omissions while remaining stable under meaning-preserving reorganizations.

\section{Discussion and Limitations}

\subsection{Discussion}

RadSEM is motivated by a practical observation: many metric failures in radiology arise not from a lack of semantic similarity in a broad sense, but from mismatched scoring units and unconstrained matching. By rewriting the Findings sections of both reports into ordered atomic finding sentences, RadSEM aligns comparison granularity with observation-level clinical correctness: individual statements about a site and its finding. This reduces sensitivity to stylistic rephrasing and section formatting, while making omissions and unsupported additions explicit as unmatched atomic finding sentences.

The second design choice, contradiction-aware many-to-many matching, directly targets clinically decisive operators such as negation, laterality, and normal/abnormal polarity while remaining tolerant of report granularity. Rather than hoping that a global similarity score will implicitly penalize contradictions, RadSEM prevents incompatible pairs from receiving credit. When independently rewritten reports express the same clinical proposition at different granularities, many-to-many pair alignment allows compatible links to be represented explicitly in the tag output. The deterministic scoring code then applies fixed pair-level coefficients, capacity-constrained matched-credit allocation, class-wise unmatched counts, the no-unmatched class rule, a no-unmatched partial adjustment rule, and class-presence-normalized final weighting. Partial credit is restricted to clinically justified cases, such as part--whole containment or incomplete abnormal-detail coverage. This design explains why RadSEM remains stable under meaning-preserving reorganization while sharply penalizing polarity flips and abnormal-to-normal conversions.

The RadSEM-Alt comparison further illustrates why RadSEM keeps many-to-many candidate links rather than forcing one-best sentence assignments during matching. The purpose of many-to-many matching is not to give unlimited credit; matched credit is still constrained later at the sentence level. Rather, candidate retention allows the metric to defer ambiguous granularity decisions until deterministic scoring. In this sense, RadSEM treats report comparison primarily as a coverage problem over two sets of atomic findings, whereas RadSEM-Alt treats it more as an early assignment problem from generated statements to reference anchors. This is important for radiology reports, where a broad generated statement may overlap with multiple atomic reference findings and where one reference finding may be expressed through several clinically distinct generated fragments. The lower SSREE consistency of RadSEM-Alt suggests that prematurely converting such ambiguity into Ref-anchored one-to-many assignments is less robust than preserving compatible links and resolving them under explicit capacity constraints. Because RadSEM-Alt changes both the alignment constraint and the subsequent unmatched-accounting policy, this comparison is best viewed as evidence for the reasonableness of RadSEM's design rather than as a causal attribution to one component.

The accepted-pair and unmatched outputs also make RadSEM interpretable beyond a scalar score. Each error can be traced to specific omitted, unsupported, contradictory, or partially matched atomic finding sentences. This property is useful for model debugging, targeted error analysis, and clinical review, especially when report-generation systems need improvement for particular finding types or anatomical sites.

These design choices reflect a broader distinction among medical imaging and multimodal model-evaluation paradigms. A platform-level framework such as MedBench v4 emphasizes scalable evaluation across heterogeneous tasks through a shared scoring interface and calibrated LLM judging \citep{ding2025medbenchv4}; GREEN emphasizes clinically informative error auditing through an expert-style LLM reviewer \citep{ostmeier2024green}; FFL-based phrasal grounding emphasizes lexicon-based finding representation and, when images and anatomical boxes are available, visual localization of finding phrases \citep{mahmood2024evaluating}; RadSEM emphasizes constrained, inspectable atomic finding sentence scoring for reference-based text comparison. RadSEM therefore shifts the language model's role from free-form judge to parser and semantic matcher. Once atomic finding sentences and match labels are fixed, the score is computed by explicit rules and can be inspected or debugged, but the framework still depends on the quality and consistency of the upstream rewriting and matching stages.

A deeper methodological issue is that radiological anatomy and report granularity do not form a single clean universal hierarchy. Organs, lobes, segments, vessels, devices, body spaces, and imaging regions may be related through part-of, located-in, adjacent-to, supplying or draining, and clinically grouped-with relations, rather than through a single tree. Moreover, a reference report is usually a clinically acceptable expression of the image rather than an exhaustive world state. RadSEM does not claim to solve this ontology problem. Instead, it treats granularity handling as a controlled engineering approximation: atomic finding sentences expose the units being compared, anti-contradiction rules prevent incompatible alignments, partial credit is limited to clinically defensible granularity gaps, and abnormal-focused weighting makes the relative importance of abnormal findings explicit. This makes the main scoring assumptions more visible than they would be in a single free-form judge prompt, while still requiring validation when the method is transferred to new modalities, body regions, languages, or reporting conventions.

FFL-based phrasal grounding also highlights an important boundary of RadSEM. When images, anatomical-region detectors, and curated chest-X-ray lexicons are available, image-grounded metrics can evaluate whether a textual finding is supported by the image rather than only whether it agrees with a reference report. Recent phrase-grounded fact-checking work pushes this idea further by explicitly detecting errors in both findings and their stated anatomical locations from image-report pairs \citep{mahmood2025phrasegrounded}. RadSEM does not perform such visual verification. Its intended use is narrower: to provide a controlled comparison between generated and reference Findings. This makes it easier to deploy across modalities and reporting styles without building a dedicated visual grounding pipeline, but it also means that RadSEM inherits the incompleteness and possible errors of the reference Findings.

\subsection{Limitations}

RadSEM currently assumes that reports can be reliably rewritten into atomic finding sentences under a consistent set of constraints. Unlike FFL-based systems that rely on curated chest-X-ray finding vocabularies and anatomical grounding resources, RadSEM relies on constrained LLM behavior during rewriting and matching. Although the released implementation uses deterministic scoring rules, including fixed partial-credit coefficients, sentence-level matched-credit capacities, explicit accounting of unmatched Ref/Gen sentences, the no-unmatched class rule, the no-unmatched partial adjustment rule, and class-presence-normalized final weighting, these rules do not by themselves guarantee that every clinical containment or granularity-compatibility decision is correct. Stage 1--2 implementations may still differ in edge cases, such as how to split borderline compound statements, how to decide whether a broad and a specific statement express the same clinical proposition or independent findings, how to interpret hedging, or how to handle highly templated structured reports. Further standardization, versioned prompts, and shared benchmark examples would improve portability.

A broader limitation concerns how radiology-report metrics should be validated. Radiologists provide indispensable clinical expertise for image interpretation and report review, but a holistic numerical rating under a vague rubric should be interpreted with attention to the scoring rubric and aggregation assumptions, rather than treated as a complete scalar ground truth for metric development. Image reading and clinical reporting are expert tasks, whereas assigning a reproducible score to a report requires additional measurement assumptions about evaluation units, class weights, omissions, unsupported findings, and partial credit. Therefore, agreement with ad hoc clinician ratings is informative but not sufficient: such ratings may vary with institutional style, the supplied rubric, and each reviewer's implicit weighting of different error types. A complementary validation strategy is to expose the metric's accepted pairs, pair labels, unmatched atomic finding sentences, scoring rules, and case-level rankings to radiologists, and ask whether the resulting error localization and score behavior are clinically useful for model comparison and debugging. In this sense, RadSEM is intended not to replace clinical judgment, but to make the quantification step explicit enough for clinical review.

Another limitation is empirical scope. The main SSREE experiment uses de-identified reports from an internal repository and controlled perturbations, and the current study does not include a prospective external-site validation or systematic radiologist adjudication of every intermediate tag. Because the de-identified reports were translated into English before evaluation, the current results also do not directly establish performance on original-language reports or on multilingual reporting conventions. Prospective external validation and systematic adjudication would be important for estimating robustness across institutions, reporting styles, languages, and clinical use cases.

A further limitation is section scope. RadSEM evaluates Findings and intentionally excludes the Impression section. This avoids using a clinically contextualized conclusion as the reference for an image-reading benchmark, but it also means that RadSEM does not measure whether a system produces the best diagnostic synthesis, communicates urgency appropriately, or recommends suitable follow-up. A future impression-level evaluator would require a different schema and should ideally receive the same clinical context available to the radiologist. For image-only generation systems, impression-like outputs should be phrased with appropriate uncertainty and should not be treated as substitutes for contextual clinical judgment.

The no-unmatched partial adjustment is intentionally mild. It is not meant to treat clinically compatible granularity differences as omissions or hallucinations. Instead, it prevents no-unmatched classes containing matched pairs with weight below 1.0 from receiving a perfect class score solely because every sentence has a compatible counterpart. This design preserves RadSEM's coverage-first behavior while making pair-level partial weights visible in the final class score. Because the adjustment is intentionally mild, it may under-penalize partial matches in settings where fine-grained abnormal detail is clinically decisive.

RadSEM also uses a small number of interpretable hyperparameters, including abnormal/normal class mixture weights and partial-credit coefficients. These values were fixed in this study, but different clinical domains, modalities, languages, or institutional reporting styles may prefer mild retuning. Finally, SSREE stresses monotonicity under controlled perturbations; complementary testing on diverse real model outputs, external datasets, and additional modalities can further identify where rewriting rules should be extended, for example to longitudinal comparison language or complex follow-up recommendations.

\section{Conclusion}

RadSEM provides a clinically oriented and interpretable metric for reference-based radiology report scoring. It builds on the broader principle of structured finding-level comparison by combining two ideas: rewriting reports into ordered atomic finding sentences and performing contradiction-aware semantic matching before scoring. This design defines stable comparison units, prevents incompatible statements from receiving undeserved credit, and yields an abnormal-focused weighted F1 that emphasizes clinically important findings. In the SSREE stress test with graded semantic corruption, RadSEM exhibits high monotonic ranking consistency and avoids common failure modes of lexical overlap metrics, especially around polarity inversions. Beyond a single scalar score, RadSEM produces accepted sentence pairs with pair labels and unmatched sentence sets that support error localization and debugging. Future work includes extending the rewriting and contradiction rules to broader modalities and languages and improving the robustness of Stage 1--2 implementations.

\clearpage
\begingroup
\footnotesize
\setlength{\bibsep}{0pt}
\bibliographystyle{unsrtnat}
\bibliography{references/reference}
\endgroup

\end{document}